\documentclass[prl,twocolumn,superscriptaddress,preprintnumbers,%
showpacs,amsmath,amssymb]{revtex4}

\usepackage{graphicx}

\newcommand{\diag}{\mathrm{diag}}
\newcommand{\etl}{\tilde{e}}
\newcommand{\Btl}{\tilde{B}}
\newcommand{\Bext}{B_{\text{ext}}}
\newcommand{\muq}{\mu_{\text{q}}}
\newcommand{\mue}{\mu_e}
\newcommand{\MeV}{\;\text{MeV}}
\newcommand{\G}{\;\text{G}}
\newcommand{\Ms}{M_s}
\newcommand{\half}{\tfrac{1}{2}}
\newcommand{\third}{\tfrac{1}{3}}
\newcommand{\twothirds}{\tfrac{2}{3}}

\begin{document}

\title{Color superconducting matter in a magnetic field}

\author{Kenji Fukushima}
\email{fuku@quark.phy.bnl.gov}
\affiliation{RIKEN BNL Research Center,
 Brookhaven National Laboratory, Upton, NY 11973, USA}
\author{Harmen J.\ Warringa}
\email{warringa@quark.phy.bnl.gov}
\affiliation{Department of Physics, Bldg.~510A,
 Brookhaven National Laboratory, Upton, NY 11973, USA}

\date{July 25, 2007}

\begin{abstract}
  We investigate the effect of a magnetic field on cold dense
  three-flavor quark matter using an effective model with four-Fermi
  interactions with electric and color neutrality taken into account.
  The gap parameters $\Delta_1$, $\Delta_2$, and $\Delta_3$
  representing respectively the predominant pairing between down and
  strange ($d$-$s$) quarks, strange and up ($s$-$u$) quarks, and up
  and down ($u$-$d$) quarks, show the de~Haas-van~Alphen effect, i.e.\
  oscillatory behavior as a function of the modified magnetic field
  $\Btl$ that can penetrate the color superconducting medium.  Without
  applying electric and color neutrality we find
  $\Delta_2\simeq\Delta_3\gg\Delta_1$ for $2\etl\Btl>\muq^2$, where
  $\etl$ is the modified electromagnetic coupling constant and $\muq$
  is one third of the baryon chemical potential.  Because the
  average Fermi surface for each pairing is affected by taking into
  account neutrality, the gap structure changes drastically in this
  case; we find $\Delta_1\gg\Delta_2\simeq\Delta_3$ for
  $2\etl\Btl>\muq^2$.  We point out that the magnetic fields as strong
  as presumably existing inside magnetars might induce significant
  deviations from the gap structure
  $\Delta_1\simeq\Delta_2\simeq\Delta_3$ at zero magnetic field.
\end{abstract}
\preprint{BNL-NT-07/33}
\maketitle


By analyzing Quantum Chromodynamics (QCD) it has been established that
at zero temperature and high enough baryon densities color
superconducting (CSC) matter should be formed~\cite{Rajagopal:2000wf}.
Unfortunately no experimental evidence for color superconductivity is
yet available.  In CSC matter Cooper pairs of quarks are created due to
an attractive interaction between quarks on opposite sides of the
Fermi surface.  The (almost) sole place where one might be able to
find color superconductivity in nature would be the central part of
neutron stars.  To this aim one has to clarify the properties of
CSC matter under the physical conditions maintained inside neutron
stars~\cite{book-shapiro,Lattimer:2006xb}.

The neutron star density is at most $\rho\!\sim\!10\rho_0$ in the
core, where $\rho_0$ is the normal nuclear density
$\sim0.17\,$nucleon/fm$^3$.  This density roughly corresponds to a
quark chemical potential (i.e.\ one third of the baryon chemical
potential) $\muq\sim500\MeV$ as deduced from $\rho\sim3\muq^3/\pi^2$.
At this intermediate density one cannot neglect the role of the
strange quark mass $\Ms=100\!\sim\!200\MeV$.  The strange quark mass
induces a ``pressure'' to tear the Cooper pairs apart, i.e., a Fermi
surface mismatch of size $\Ms^2/2\muq$ between $u$/$d$ quarks and $s$
quarks will be formed.  The pairing pattern is quite complicated in
the density region where $\Ms^2/2\muq$ is comparable to the gap energy
$\Delta$ which are both of order tens MeV around the region of our
interest where $\muq\sim500\MeV$.

The Fermi surface mismatch is further caused by the requirement of
neutrality which is broken by $\Ms\neq0$ in three-flavor quark matter.
The system should be electric neutral to avoid divergent field
energies faster than the volume, otherwise the system is not stable
thermodynamically.  Regarding color neutrality the constraint is more
stringent, that is, the whole system must be a color-singlet.  To
consider the phase structure, however, it is adequate to impose
global color neutrality as well as global electric neutrality.  In
the effective model we will use in this Letter this can be achieved by
introducing the electric and color chemical potentials, $\mue$,
$\mu_3$, and $\mu_8$ corresponding the the \textit{negative} electric
charge matrix $Q_e=-Q=\diag(-\twothirds,\third,\third)$ in flavor
space and two diagonal color charge matrices
$T_3=\diag(\half,-\half,0)$ and $T_8=\diag(\third,\third,-\twothirds)$
in color space~\cite{Iida:2000ha}.  These chemical potentials mimic
gauge field dynamics.

In neutron stars another source for the ``pressure'' on Cooper pairs
is a magnetic field.  In regular neutron stars the magnetic field
strength on the surface is of order $B=10^9\!\sim\!10^{12}\G$ and it
reaches values as large as $B\sim10^{15}\G$ (in which case
$eB=1\!\sim\!10\MeV^2$) in a special kind of neutron star called
magnetar~\cite{Duncan:1992hi}).  Actually the virial
theorem~\cite{lai-shapiro} enables us to deduce $B \lesssim 10^{18}\G$
in the interior of the neutron star.  In compact stars that are
self-bound rather than gravitationally bound, the maximum magnetic
field could be even larger such that $e B \approx \muq^2$.   Clearly,
the effect of these magnetic fields are not to be neglected at all as
compared to $\Delta$ and $\Ms^2/2\muq$.

Much work has been done to investigate the effect of the magnetic
field on nuclear matter (see Ref.~\cite{Lattimer:2006xb} and
references therein) and also some on normal quark
matter~\cite{Gusynin:1994re,Ebert:1999ht} but only limited
results~\cite{Alford:1999pb,Ferrer:2005vd,Manuel:2005hu,Ferrer:2007iw}
are available on CSC matter in a penetrating magnetic field.

In particular in Ref.~\cite{Ferrer:2005vd} the analytical solution of
the gap equation is found only in the limit of a strong magnetic field
without taking account of the neutrality conditions.  The new material
we shall elucidate in this Letter is twofold; first, we solve the
gap equations numerically in order to get the gap parameters for any
value of magnetic field within the framework of the
Nambu--Jona-Lasinio (NJL) model. This model correctly describes qualitative
features of high-density QCD~\cite{Rajagopal:2000wf,Buballa:2003qv}.
Next, we impose the electric and color neutrality conditions on the
system which changes the qualitative behavior of the gap parameters
significantly from the non-neutral case.


We assume only the predominant pairing in the color anti-symmetric
channel;
\begin{equation}
 \langle\bar{\psi}_{i\alpha}\gamma_5 C\bar{\psi}_{j\beta}^T\rangle
  \!\sim\! \epsilon_{1\alpha\beta}\epsilon_{1ij}\Delta_1
  \!+ \epsilon_{2\alpha\beta}\epsilon_{2ij}\Delta_2
  \!+ \epsilon_{3\alpha\beta}\epsilon_{3ij}\Delta_3 \,,
\end{equation}
where $\alpha$ and $\beta$ run from 1 to 3 in color space
($r$,$g$,$b$) and $i$ and $j$ run from 1 to 3 in flavor space
($u$,$d$,$s$).  The gap parameters $\Delta_1$, $\Delta_2$, and
$\Delta_3$ represent the $ds$-pairing, $su$-pairing, and $ud$-pairing,
respectively.

This gap pattern breaks the electromagnetic $\mathrm{U}(1)$ symmetry.
As a result the photon becomes massive, so that a pure electromagnetic
field cannot penetrate CSC matter.  However, the pairing pattern is
still invariant under $\mathrm{U}(1)_{\tilde{Q}}$
transformations, where $\tilde{Q}=\boldsymbol{1}_{\text{color}}\otimes
Q - Q\otimes\boldsymbol{1}_{\text{flavor}}$.  The corresponding
rotated electromagnetic field is
$\tilde{A}_\mu = A_\mu\cos\theta - G_\mu^Q\sin\theta$, which is a
combination of the electromagnetic field $A_\mu$ and a component of
the gluonic field $G_\mu^Q$.  The rotated photon stays massless, hence
a rotated magnetic field $\Btl$ can penetrate CSC
matter~\cite{Alford:1999pb}.  The coupling constant for $\tilde{A}$ is
$\etl =e\cos\theta$ where $e$ and $g$ are the electromagnetic and QCD
coupling constant.   Here the mixing angle $\theta$ may depend on the
gap structure~\cite{Alford:1999pb,Gorbar:2000ms}.  In the convention
that the gauge fields are defined with generators normalized as
$\mathrm{tr}[t_a t_b]=2\delta_{ab}$, one finds
$\cos\theta=g/(\third e^2+ g^2)^{1/2}$ for color-flavor locked (CFL)
matter ($\Delta_1 \approx \Delta_2 \approx \Delta_3$).  For the phases
with only $\Delta_2$ or $\Delta_3$ nonzero (i.e.\ 2SCsu or 2SC phase)
one has $\cos\theta = g /(\tfrac{1}{12} e^2+ g^2)^{1/2}$.
Interestingly enough, we find that the mixing angle in the phase with
only $\Delta_1$ nonzero (i.e.\ 2SCds phase) is the same as that in the
CFL phase~\cite{future}.

By careful consideration of the boundary layer between CSC matter and
normal quark matter one can derive a relation between the magnitude
of the applied external magnetic field $\Bext$ outside and the rotated
magnetic field $\Btl$ inside CSC matter. For sharp boundaries
(boundary smaller than screening length) a small part of the flux is
expelled and one has $\Btl\approx\Bext\cos\theta$, while for smooth
boundaries $\Btl \approx \Bext$ hence no flux is expelled at
all.~\cite{Alford:1999pb}.  Since $g \gg e$ in the region we are
interested in, $\cos \theta \approx 1$ which implies that the
magnitudes of the magnetic fields outside and inside the CSC core are
approximately equal, that is, we will implicitly assume
$\etl\Btl\approx e\Bext$ in our discussions.

While the quark Cooper pairs are neutral with respect to the
$\tilde{Q}$ charge, some of the individual quarks which form a pair
are, however, charged under the $\Btl$ field.  In particular, the
three flavors and three colors result in nine different quarks from
which four are $\tilde{Q}$ charged.  We summarize in the following
table the quark species, $\tilde{Q}$, and the gap parameters involved
in the pairing;
\begin{center}
 \begin{tabular}{|c||c|c|c||c|c||c|c||c|c||}
 \hline
  pairing & $ru$ & $gd$ & $bs$  & $bd$ & $gs$ & $rs$ & $bu$
          & $gu$ & $rd$ \\
 \hline
  $\tilde{Q}$ & 0 & 0 & 0 & 0 & 0 & $+1$ & $-1$ & $-1$ & $+1$ \\
 \hline
  gap & \multicolumn{3}{c||}{$\Delta_1$ $\Delta_2$ $\Delta_3$}
      & \multicolumn{2}{c||}{$\Delta_1$}
      & \multicolumn{2}{c||}{$\Delta_2$}
      & \multicolumn{2}{c||}{$\Delta_3$}\\ \hline
 \end{tabular}
\end{center}

For the charged sectors a constant magnetic field results in the
Landau quantization of the energy dispersion relation.  For instance,
the $rs$-$bu$ sector of the Nambu-Gor'kov propagator contains sixteen
dispersion relations as a function of $p^2=p_x^2+p_y^2+p_z^2$.  Once
we turn on a constant $\Btl$ in the $z$-direction, rotational symmetry
is broken and the dispersion relations are modified by the following
replacement,
\begin{equation}
  p^2 \to p_B^2 = 2\vert\etl\Btl\vert (n+\half) \pm
   \vert\etl\Btl\vert + p_z^2 \,,
\label{eq:replace}
\end{equation} 
where $n=0,2,3\dots$ with $\pm\vert\etl\Btl\vert$ depending on the
spin and $\tilde Q$ charge.  Then, in order to write down the
thermodynamic potential $\Omega$ of the NJL model, we can utilize the
conventional expression (see e.g.\  Ref.~\cite{Fukushima:2004zq}
including full account of $\Ms\neq0$) with the prescription
(\ref{eq:replace}) for the $rs$-$bu$ and $gu$-$rd$ sectors. The
associated momentum integration is modified as follows
\begin{equation}
 \int\frac{dp_x dp_y}{(2\pi)^2} \to \frac{\vert \etl\Btl \vert}{2\pi}
  \sum_{n=0}^{\infty} \,.
\end{equation}
We will present the prove of this simple replacement in
a separate paper~\cite{future}.


\begin{figure}
 \includegraphics[width=8cm]{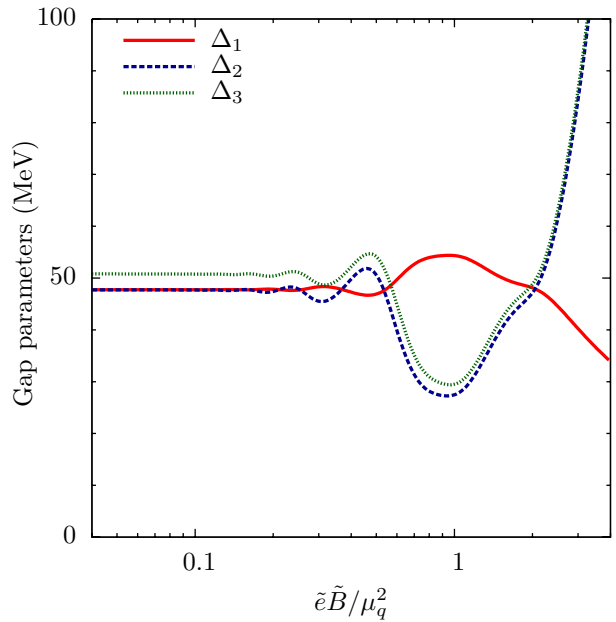}
 \caption{Gap parameters as a function of $\etl\Btl/\muq^2$ for
  $\muq=500\MeV$ without neutrality.}
 \label{fig:nn00}
\end{figure}

We first solve three gap equations without neutrality taken into
account,
\begin{equation}
 \frac{\partial\Omega}{\partial\Delta_1}
 =\frac{\partial\Omega}{\partial\Delta_2}
 =\frac{\partial\Omega}{\partial\Delta_3}=0
\end{equation}
at $\mue=\mu_3=\mu_8=0$ to check consistency with
Ref.~\cite{Ferrer:2005vd}.

We show our numerical results as a function of a dimensionless
parameter $\etl\Btl/\muq^2$ in Fig.~\ref{fig:nn00} for
$\muq=500\MeV$.  We chose the cut-off parameter $\Lambda=1000\MeV$
and the four-Fermi coupling constant to yield $\Delta_0=50\MeV$ for
$\Ms=\Btl=0$.  We made sure that $\Lambda$-dependence is tiny once the
coupling constant runs as a function of $\Lambda$ to give a fixed
value of $\Delta_0$.  To reduce the cut-off artifact, we used a
smooth Fermi-Dirac-like form factor,
$\half\{1-\tanh[(p_B-\Lambda)/\omega]\}$ with a choice
$\omega=0.1\Lambda$ in the momentum integration.  If we adopt a
smaller value of $\omega$ which results in a sharper cut-off scheme,
the curves in Fig.~\ref{fig:nn00} become less smooth.  In particular,
for $\omega\to 0$ we find tiny spikes in the gap parameters when
$\Lambda^2/(2\etl\Btl)$ is an integer, originating from the vacuum
energy contribution to the thermodynamic potential.  We have checked
the robustness of the smooth oscillatory shapes seen in
Fig.~\ref{fig:nn00} by varying the value of
$\omega$~\cite{future}.

We can see that $\Delta_2$ and $\Delta_3$ are close to each other
apart from a discrepancy by the strange quark mass which is
$\Ms=100\MeV$ in our calculation; the effect of $\Ms$ pushes
$\Delta_1$ and $\Delta_2$ down to $47.7\MeV$ and $\Delta_3$ up to
$50.8\MeV$ at $\Btl=0$.  The gap parameters show oscillatory behavior
as long as $2\etl\Btl<\muq^2$ because the density of states increases
every time $2\etl\Btl n$ approaches $\muq^2$ for some non-zero $n$.
Thus, the oscillation ceases when the first Landau level ($n=1$) lies
above the Fermi surface, i.e.\ $2\etl\Btl>\muq^2$.  In fact, it is
manifest in Fig.~\ref{fig:nn00}.  The behavior of the gap parameters
is similar to the oscillation of the magnetization of a material in an
external magnetic field which is known as the de~Haas-van~Alphen
effect.

Our numerical results are qualitatively consistent with the analytical
evaluation in Ref.~\cite{Ferrer:2005vd} for $\etl\Btl\gg\muq^2$ when
only the lowest Landau level (LLL) contributes to the gap equations.
In that case the gap parameters $\Delta_2$ and $\Delta_3$ increase
monotonically as a function of $\Btl$.  The reason for this is
understood in view of the analytical expressions given by Eqs.~(95)
and (96) in the second paper of Ref.~\cite{Ferrer:2005vd}; roughly
speaking, as a result of the LLL approximation, the phase space is
enlarged as $2\muq^2\to\muq^2+\etl\Btl$.  At the same time we see that
$\Delta_1$ decreases slightly as a function of $\Btl$ because it is
only indirectly sensitive to $\Btl$, which is qualitatively consistent
with Eq.~(101) in Ref.~\cite{Ferrer:2005vd}.

\begin{figure}
 \includegraphics[width=8cm]{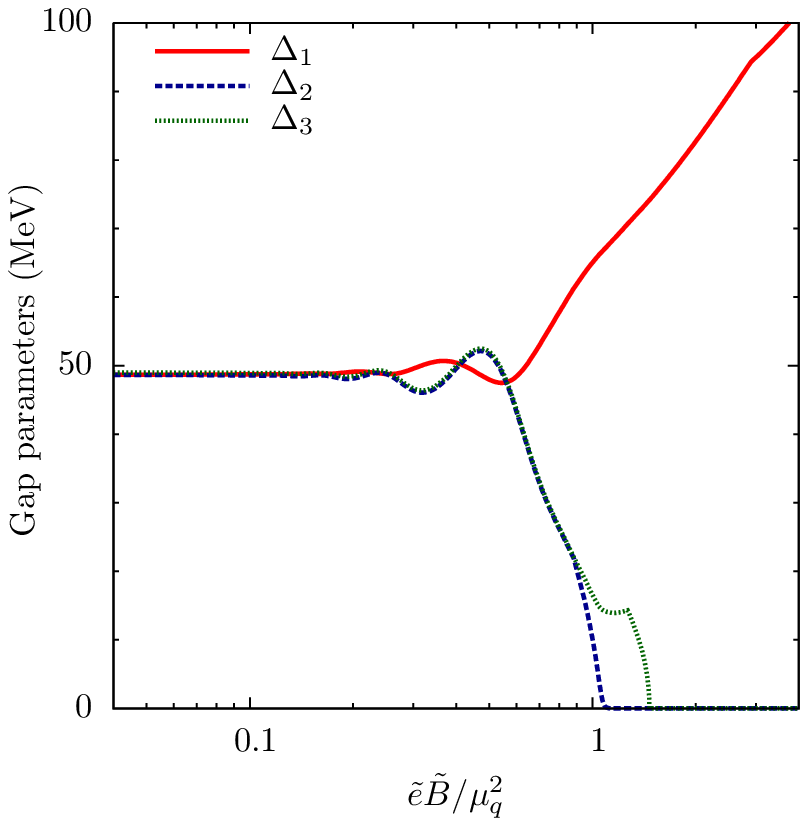}
 \caption{Gap parameters as a function of $\etl\Btl/\muq^2$ with
 neutrality, for $\muq = 500 \;\mathrm{MeV}$.}
 \label{fig:t00}
\end{figure}

\begin{figure}
 \includegraphics[width=8cm]{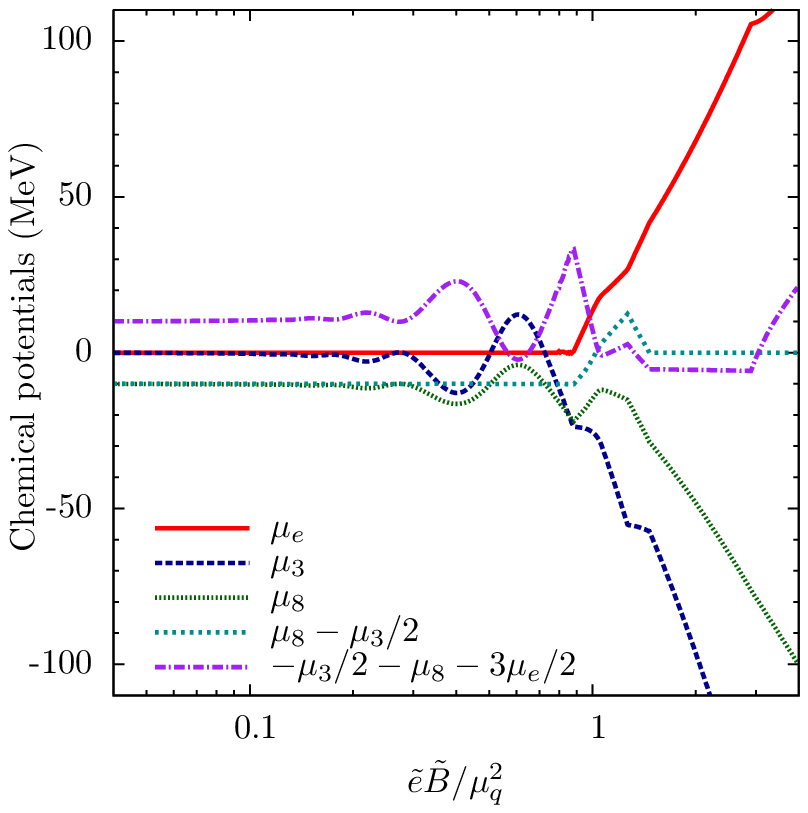}
 \caption{Chemical potentials as a function of $\etl\Btl / \muq^2$,
 for $\muq = 500\;\mathrm{MeV}$.}
 \label{fig:ct00}
\end{figure}

Next we will impose three neutrality conditions,
\begin{equation}
  \frac{\partial\Omega}{\partial\mue}
 =\frac{\partial\Omega}{\partial\mu_3}
 =\frac{\partial\Omega}{\partial\mu_8}=0 \,.
\end{equation}
We have also added the contribution of a free electron and muon gas to
$\Omega$.  Electrons and muons feel the magnetic field
$\Btl\cos\theta$ with the coupling constant $e$, that amounts to
$e\Btl\cos\theta=\etl\Btl$.  Once we take account of neutrality the
situation drastically changes.  Figure~\ref{fig:t00} shows the gap
parameters with $\mue$, $\mu_3$, and $\mu_8$ determined
self-consistently for $\muq=500\MeV$.  The corresponding chemical
potentials are displayed in Fig.~\ref{fig:ct00}.  It can be seen in
Fig.~\ref{fig:ct00} that at very large $\etl\Btl$, $\mue$ is larger
than the muon mass $\simeq 106\MeV$.  This indicates that in that case
a significant number of muons will be present in the system.

In sharp contrast to the non-neutral case, we find that $\Delta_1$
grows with increasing $\etl\Btl$. The gap parameters $\Delta_2$ and
$\Delta_3$ vanish smoothly at $\etl\Btl/\muq^2\simeq 1.09$ and $1.47$,
respectively.  This implies a second order transition from the CFL
phase to the so-called dSC phase, followed by a transition to the
so-called 2SCds phase.  Note that the mixing angle $\theta$ is common
in the CFL, dSC, and 2SCds phases.  The behavior of $\Delta_1$ at
large $\etl\Btl$ can be accounted for by $\mu_3$ and $\mu_8$; the
Fermi surface average $\bar{\mu}_{\text{$bd$-$gs$}}$ becomes larger,
for example in our calculation, from $500\MeV$ at $\Btl=0$ to
$625\MeV$ at $\etl\Btl/\muq^2=4$.  This results in a larger gap
parameter.  So the system exhibits a phase with only $\Delta_1$
nonzero at large $\etl\Btl/\muq^2$.  This implies two-flavor color
superconducting pairing between $d$ and $s$ quarks, hence the name
2SCds.  The possibility of the 2SCds phase as a ground state for
$\etl\Btl/\muq^2>1.47$ is quite interesting, since the 2SCds phase has
rarely been paid attention to in the QCD phase diagram.  [See
Ref.~\cite{Warringa:2005jh} for detailed analyses including the 2SCds
region.]

The 2SCds phase is similar to the more familiar 2SC phase where only
$\Delta_3$ takes a finite value.  The behavior of $\mu_3$ and $\mu_8$
in the 2SCds region might look totally different from the 2SC phase in
which $\mu_3=0$ and $\mu_8\ll\mue$.  This difference, however, turns
out superficial once we rearrange the color-flavor bases properly as
$\mu_3'=\mu_8-\half\mu_3$ and
$\mu_8'=-\half\mu_3-\mu_8-\tfrac{3}{2}\mue$.  Then, as shown by dotted
curves in Fig.~\ref{fig:ct00}, $\mu_3'$ in the 2SCds phase is zero
just like $\mu_3$ in the 2SC phase, and $\mu_8'$ stays smaller than
$\mue$ just like $\mu_8$ in the 2SC phase.  We note that $\mu_3'$ is
not oscillatory at all even though $\mu_3$ and $\mu_8$ are.  We also
point out that CFL matter remains a $\tilde{Q}$-insulator, i.e.\
$\mue=0$~\cite{Rajagopal:2000ff} until $\etl\Btl/\muq^2\simeq0.88$.

In fact, there are two windows in which gapless dispersion relations
with non-zero gap parameters appear;
$0.88 < \etl\Btl/\muq^2 < 1.09$ for $rs$-$bu$ pairing with
$\Delta_2$ and $1.25 < \etl\Btl/\muq^2 < 1.47$ for $gu$-$rd$ pairing
with $\Delta_3$. The 2SCds phase for $\etl\Btl/\muq^2>1.47$ is fully
gapped.  Once the system enters the gapless state,
$\Delta_2$ and $\Delta_3$ rapidly decrease to zero.  Because the phase
space is enlarged by large $\Btl$ for $\tilde{Q}$-charged quarks and
thus their density increases, the Fermi surface averages
$\bar{\mu}_{\text{$rs$-$bu$}}$ and $\bar{\mu}_{\text{$gu$-$rd$}}$
should be located lower than the others to keep neutrality.  It can
happen with developing $\mu_3\neq0$ and $\mu_8\neq0$ which alter not
only the Fermi surface averages but also the Fermi surface mismatches
$\delta\mu_{\text{$rs$-$bu$}}$ and $\delta\mu_{\text{$gu$-$rd$}}$.
Under the constraint of neutrality, hence, $\Delta_2$ and $\Delta_3$
become smaller because of the decreasing Fermi surface average with
increasing $\Btl$, and at the same time, the Fermi surface mismatches
for $\Delta_2$ and $\Delta_3$ grow up with $\mu_3$ and $\mu_8$ induced
by $\Btl$, and eventually the gapless dispersion relations emerge when
$\delta\mu_{\text{$rs$-$bu$}}>\Delta_2$ or
$\delta\mu_{\text{$gu$-$rd$}}>\Delta_3$.

The question of whether the 2SCds phase is a real possibility of the
ground state under a sufficiently strong magnetic field or not should
be answered by energy comparison with normal quark matter.  We have to
calculate the energy density in CSC matter with $\etl\Btl$ and that in
normal quark matter with $e\Bext$ to determine which is energetically
favored.  Unfortunately the energy density has a huge oscillation as a
function of $\Btl$ if $\etl\Btl \simeq \Lambda^2$.  This oscillation
arises from the vacuum energy contributions to the thermodynamic
potential.  Because normal quark matter couples in another way to the
magnetic field, the oscillation in the energy density of normal quark
matter is different from that of CSC matter, making energy comparison
at $\etl\Btl \simeq \Lambda^2$ ambiguous.

We have carefully checked that the cut-off dependence is tiny in the
gap equations and neutrality conditions as we mentioned, and it is
natural because the momentum integration near the Fermi surface should
be dominant. Therefore we could as well have taken a larger value of
$\Lambda$ such that the gap parameters remain to have the same value
and $\etl\Btl$ is much smaller than $\Lambda^2$.  In this way we found
that there is always an energy gain between CSC and normal quark matter,
in particular it is kept to be at least $2\times10^8\MeV^4$ around 
$\etl\Btl/\muq^2\sim 1$.

However the field energy it takes to expel part of the applied
magnetic field from CSC matter is in the case of a sharp boundary equal
to $\half(\Bext^2-\cos^2\theta\Bext^2)\lesssim e^2\Bext^2/6g^2$.  The
coefficient $e^2/ g^2$ is of order $0.01$, and therefore, the energy
cost is of order $1 \times 10^8 \MeV^4$ around $\etl\Btl/\muq^2\sim
1$, which is comparable or less than the energy gain.  In the case of
a smooth boundary no flux is expelled at all, so then CSC matter is
always favored.  We might thus expect to see
$\Delta_1\gg\Delta_2\simeq\Delta_3$ before reaching a possible
transition to normal quark matter.  Hence if
$e \Bext \approx \muq^2$, the 2SCds phase could be a likely candidate
for the ground state of matter inside magnetars.

     In summary, we considered the effect of a strong magnetic field
on neutral CSC quark matter.  We found that the neutrality conditions
significantly change the non-neutral results.  We pointed out the
possibility of the 2SCds phase in the interior of the magnetar.  

     K.~F.\ thanks T.~Kunihiro and M.~Tachibana for comments.  This
research was supported in part by RIKEN BNL Research Center and the
U.S.\ Department of Energy under cooperative research agreement
\#DE-AC02-98CH10886.

\end{document}